\documentclass[conference,twocolumn]{IEEEtran}

\usepackage{graphicx}
\usepackage{color}
\usepackage{placeins}
\usepackage{float}
\usepackage{tabularx,colortbl}
\usepackage{amssymb}
\usepackage{amsthm}
\usepackage{cite}
\usepackage{amsmath}

\usepackage{caption2}
\theoremstyle{plain}
\newtheorem{thm}{Theorem}

\theoremstyle{plain}

\newtheorem{cor}{Corollary}

\IEEEoverridecommandlockouts

\begin{document}

\title{Uplink Performance of High-Mobility Cell-Free Massive MIMO-OFDM Systems}

\author{Jiakang Zheng, Jiayi Zhang, Enyu Shi, Jing Jiang, and Bo Ai

\thanks{J. Zheng, J. Zhang, and E. Shi are with the School of Electronics and Information Engineering, Beijing Jiaotong University, Beijing 100044, P. R. China (e-mail: \{20111047, jiayizhang\}@bjtu.edu.cn).}
\thanks{Jing Jiang is with the Shaanxi Key Laboratory of Information Communication Network and Security, Xi'an University of Posts and Telecommunications, Xi'an 710121, China (e-mail:jiangjing@xupt.edu.cn).}
\thanks{B. Ai is with the State Key Laboratory of Rail Traffic Control and Safety, Beijing Jiaotong University, Beijing 100044, China (e-mail: boai@bjtu.edu.cn).}
}

\maketitle
\vspace{-1cm}
\begin{abstract}
High-speed train (HST) communications with orthogonal frequency division multiplexing (OFDM) techniques have received significant attention in recent years. Besides, cell-free (CF) massive multiple-input multiple-output (MIMO) is considered a promising technology to achieve the ultimate performance limit. In this paper, we focus on the performance of CF massive MIMO-OFDM systems with both matched filter and large-scale fading decoding (LSFD) receivers in HST communications. HST communications with small cell and cellular massive MIMO-OFDM systems are also analyzed for comparison. Considering the bad effect of Doppler frequency offset (DFO) on system performance, exact closed-form expressions for uplink spectral efficiency (SE) of all systems are derived. According to the simulation results, we find that the CF massive MIMO-OFDM system with LSFD achieves both larger SE and lower SE drop percentages than other systems. In addition, increasing the number of access points (APs) and antennas per AP can effectively compensate for the performance loss from the DFO.
Moreover, there is an optimal vertical distance between APs and HST to achieve the maximum SE.
\end{abstract}


\IEEEpeerreviewmaketitle

\section{Introduction}

Cell-free (CF) massive multiple-input multiple-output (MIMO) has been recently proposed as a promising physical-layer technology for achieving high and uniform spectral efficiency (SE) in wireless networks \cite{zhang2020prospective,Ngo2017Cell}. CF massive MIMO systems consist of a large number of geographically distributed access points (APs) connected to a central processing unit (CPU), and  coherently serve all user equipments by spatial multiplexing on the same time-frequency resource \cite{Ngo2017Cell}. Moreover, CF massive MIMO systems can be deployed to guarantee good coverage without cells or cell edges \cite{bjornson2019making}. The results in \cite{bjornson2019making} revealed that the CF massive MIMO system achieves better performance than small cell and cellular massive MIMO systems in terms of 95\%-likely per-user SE. Therefore, a large amount of fundamental and important aspects of CF massive MIMO have been investigated in recent years. For instance, it was shown in \cite{9322468} that, compared with the matched filter (MF) receiver, CF massive MIMO systems with the large-scale fading decoding (LSFD) receiver can achieve two-fold gains. The reason is that the fading statistics information of the entire network is utilized to calculate the LSFD weight coefficient and hence reduce interference.

Railway communication has attracted significant attention from both academia and industry due to the booming development of railways, especially high-speed train (HST) communications \cite{ai20205g}. Viaducts and tunnels are the two typical scenarios in wireless propagation environment for HST \cite{6469255}. Specifically, 86.5\% of railways is elevated in the Beijing-Shanghai HST \cite{7155729}. Therefore, there are few multi-path because of little scattering and reflection, and the line-of-sight (LoS) assumption is widely used in HST communications. For example, assuming small-scale fading is a constant value, \cite{7842154} studied the performance limits of uplink wireless delay-limited information transmission in HSTs under two trains encountering scenario, and \cite{7898862} analyzed the beamforming design principles for HST communications based on the location information.

In addition, the orthogonal frequency-division multiplexing (OFDM) technique is applied to the long term evolution for railway (LTE-R) for providing a seamless connection to existing ground cellular network \cite{7921554,7553613}. However, in high mobility scenarios, Doppler frequency offset (DFO), phase noise and timing offset can trigger the inter-carrier interference (ICI) and seriously decrease the performance of HST systems \cite{6469255,6189004}. Moreover, fast signal processing in light of high mobility and the frequency handover between adjacent base stations (BSs) are two of the main challenges in HST system design \cite{6180090}. One of the promising solutions to tackle these challenges is to utilize a distributed massive MIMO architecture named CF massive MIMO \cite{8768014}. Authors in \cite{gao2020uplink} investigated the crowded CF massive MIMO-OFDM system with spatially and frequently correlated channels, but the important ICI problem is ignored. Therefore, the fundamental limits of HST communications with CF massive MIMO-OFDM systems is still an open question.

Motivated by the aforementioned observation, we investigate the performance of the CF massive MIMO-OFDM system in HST communications, where high mobility may destroy the orthogonality of subcarriers to cause serious ICI. Considering both MF and LSFD receivers, we derive closed-form expressions for uplink SE to quantify the DFO effect. Moreover, the performance of HST communications with conventional small cell and cellular massive MIMO-OFDM systems are analyzed for comparison. Our results show that, in HST communications, the CF massive MIMO-OFDM system with LSFD receiver performs better than conventional systems.

\textit{Notation:} We use boldface lowercase letters $\mathbf{x}$ and boldface uppercase letters $\mathbf{X}$ to represent column vectors and matrices, respectively.
The $n\times n$ identity matrix is ${{\mathbf{I}}_n}$.
Superscripts $x^\mathrm{*}$, $\mathbf{x}^\mathrm{T}$ and $\mathbf{x}^\mathrm{H}$ are used to denote conjugate, transpose and conjugate transpose, respectively.
The absolute value, the Euclidean norm, the trace operator and the definitions are denoted by $\left|  \cdot  \right|$, $\left\|  \cdot  \right\|$, ${\text{tr}}\left(  \cdot  \right)$, and $\triangleq$, respectively.
Finally, $x \sim \mathcal{C}\mathcal{N}\left( {0,{\sigma^2}} \right)$ represents a circularly symmetric complex Gaussian random variable $x$ with variance $\sigma^2$.


\section{System Model}\label{se:model}

\begin{figure}[t]
\vspace{0.2cm}
\centering
\includegraphics[scale=0.6]{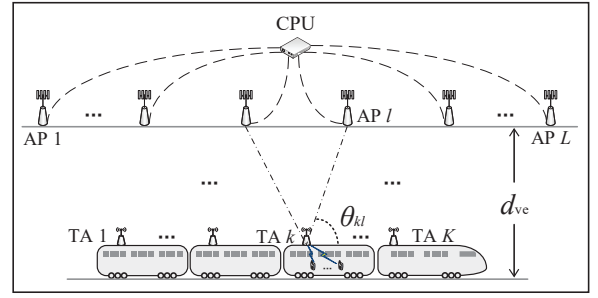}
\caption{HST with CF massive MIMO-OFDM systems.} \vspace{-4mm}
\label{HST}
\end{figure}

As illustrated in Fig.~\ref{HST}, we consider a HST communication with CF massive MIMO-OFDM systems consisting of $L$ APs and $K$ mobile single-antenna train antennas (TAs) on the top of a HST. Each AP is equipped with $N$ antennas. The APs are connected to a CPU via fronthaul links. We assume that, on the same time-frequency resource, all $L$ APs serve all $K$ TAs at the same time. The arrival of angle (AOA) between AP $l$ and TA $k$ is $\theta_{kl}$.

\subsection{Propagation Model}

A 2D plane coordinate system is used to determine the locations of $L$ APs and $K$ TAs. We assume the coordinates of APs are ${{\mathbf{q}}_l} \triangleq \left[ {{a_l},{d_\text{ve}}} \right],l = 1, \ldots ,L$, where ${a_l}$ is the horizontal coordinate of AP and ${d_\text{ve}}$ is the vertical distance between the train and APs. Moreover, the coordinates of TAs are ${{\mathbf{q}}_k}\left[ n \right] \triangleq \left[ {{a_k}\left[ n \right] ,0} \right],k = 1, \ldots ,K$. Then, the abscissa position of TA $k$ at the $n$th time interval can be expressed as
\begin{align}
{a_k}\left[ n \right] = {a_k}\left[ 0 \right] + dn, n =  \ldots , - 1, 0, 1, \ldots ,
\end{align}
where ${a_k}\left[ 0 \right]$ is the initial reference abscissa position of TA $k$, and $d$ is the distance that the train travels in a time interval. Note that $n$ can be positive or negative, which respectively represents the forward and backward of the HST. Furthermore, we can obtain the straight line distance between AP $l$ and TA $k$ as $d_{kl}\left[ n \right] = \left\| {{{\mathbf{q}}_l} - {{\mathbf{q}}_k}\left[ n \right]} \right\|$, and abscissa difference between them is $d_{kl}^{{\text{ho}}}\left[ n \right] = {a_l} - {a_k}\left[ n \right]$. Then, the cosine value of the AOA between AP $l$ and TA $k$ can be given by
\begin{align}
\cos \left( {{\theta _{kl}}\left[ n \right]} \right) = \frac{{d_{kl}^{{\text{ho}}}\left[ n \right]}}{{{d_{kl}}\left[ n \right]}}.
\end{align}
Due to the high probability of LoS link in HST communications, the large-scale
fading between the AP $l$ and TA $k$ can be modeled by ${\beta _{kl}}\left[ n \right] \!=\! {\left( {{d_{kl}}\left[ n \right]} \right)^{ - \alpha }}$, where $\alpha  \!\in\! \left[ {2:6} \right]$. Then, we can obtain the channel gain between AP $l$ and TA $k$ as
\begin{align}\label{h_kl}
  {{\mathbf{h}}_{kl}}\left[ n \right] &= \sqrt {{\beta _{kl}}\left[ n \right]} \left[ {1\;\exp \left( {j2\pi {d_\text{H}}\sin \left( {{\varphi _{kl}}\left[ n \right]} \right)} \right) \ldots } \right. \notag \\
  &{\left. {\exp \left( {j2\pi {d_\text{H}}(N - 1)\sin \left( {{\varphi _{kl}}\left[ n \right]} \right)} \right)} \right]^\text{T}} ,
\end{align}
where $d_\text{H} \leqslant 0.5$ is the antenna spacing parameter (in fractions of the wavelength). Note that ${{\varphi _{kl}}\left[ n \right]}$ is the azimuth angle to the TA, and we have $\sin \left( {{\varphi _{kl}}\left[ n \right]} \right) = \cos \left( {{\theta _{kl}}\left[ n \right]} \right)$.

\subsection{Time-Domain Expression}

When the HST travels along the railway, the signal transmitted from the $k$th TA to the $l$th AP is modeled as
\begin{align}
{{\mathbf{y}}_{kl}}\left[ n \right] = \sqrt {{p_k}} {{\mathbf{h}}_{kl}}\left[ n \right]{x_k}\left[ {n - {\tau }} \right] + {{\mathbf{w}}_l}\left[ n \right],
\end{align}
where ${x_k}\left[ {n - {\tau }} \right]$ is the normalized transmit signal from TA $k$, $p_k$ denotes the uplink transmission power of TA $k$, and $\tau$ is the time-delay of propagation path, which is less than the guard interval. In addition, ${\mathbf{y}}_{kl}\left[ n \right]$, ${{\mathbf{h}}_{kl}}\left[ n \right]$ and ${{\mathbf{w}}_l}\left[ n \right]$ are the time domain received signal, the unitary time domain channel fading vector and the circular complex white Gaussian noise, respectively. Then, adding CFO to each path, the total received signal at AP $l$ can be expressed as
\begin{align}\label{time}
{{\mathbf{y}}_l}\left[ n \right] &= \sum\limits_{i = 1}^K {\sqrt {{p_i}} {\exp \left({jw \cos  \left( {{\theta _{il}}} \right) n}\right)}} \notag\\
&\;\;\;\;\;\;\;\;\;\;\;\times{{\mathbf{h}}_{il}}\left[ n \right]{x_i}\left[ {n - {\tau _{il}}} \right] + {{\mathbf{w}}_l}\left[ n \right],
\end{align}
where $w = \frac{{fvT}}{c}$ is the maximum normalized DFO. Besides, $v$ and $c$ are the velocity of train and light, respectively. Furthermore, $f$ and $T$ denote the carrier frequency and the time domain sampling duration of OFDM signals. It should be noted that $\tau_{il}$ is the time-delay of each path and the largest $\tau_{il}$ is assumed to be less than the guard interval. Therefore, the time-delay of each path can be ignored. This is because the distance difference among the paths are relatively short, comparing to the distance of electromagnetic wave propagation in one sampling duration.

\subsection{Frequency-Domain Expression}

By doing fourier transformation on the received signal \eqref{time} in time domain, the demodulated signal at AP $l$ can be given in the frequency domain as
\begin{align}\label{frequency}
{{\mathbf{y}}_l}\left[ s \right] \!=\! \sum\limits_{i = 1}^K {\sum\limits_{m = 1}^M {\sqrt {{p_i}} } {I_{il}}\left[ {m \!-\! s} \right]{{\mathbf{h}}_{il}}\left[ m \right]{x_i}} \left[ m \right] \!+\! {{\mathbf{w}}_l}\left[ s \right],
\end{align}
where ${{\mathbf{h}}_{il}}\left[ m \right]$ and ${x_i} \left[ m \right]$ are channel fading vector and transmitted signal at the $m$th subcarrier $\left( {m = 1,2, \ldots ,M} \right)$, respectively. Moreover, ${{\mathbf{w}}_l}\left[ s \right] \sim \mathcal{C}\mathcal{N}\left( {{\mathbf{0}},{\sigma ^2}{{\mathbf{I}}_N}} \right)$ is the receiver noise at the $s$th subcarrier, $M$ is the total subcarrier number of OFDM system, and ${I_{il}}\left[ {m \!-\! s} \right]$ is ICI coefficient between the $m$th and $s$th subcarriers, which can be expressed as
\begin{align}\label{I_kl}
{I_{il}}\left[ {m - s} \right] &= \frac{{\sin \left( {\pi \left( {m + {\varepsilon _{il}} - s} \right)} \right)}}{{M\sin \left( {\frac{\pi }{M}\left( {m + {\varepsilon _{il}} - s} \right)} \right)}} \notag\\
&\times \exp \left( {j\pi \left( {1 - \frac{1}{{\;M}}} \right)\left( {m + {\varepsilon _{il}} - s} \right)} \right),
\end{align}
where ${\varepsilon _{il}} = w\cos \left( {{\theta _{il}}} \right)$ is the normalized DFO between the $l$th receive AP and the $i$th transmit TA.

\section{Performance Analysis}\label{se:performance}

In this section, we study the uplink SE of CF massive MIMO-OFDM systems in HST communications. Meanwhile, small cell and cellular massive MIMO-OFDM systems are analyzed for comparison. We also derive closed-form expressions for all considered systems.

\subsection{CF Massive MIMO-OFDM}

In order to detect the symbol from the $k$th TA at $s$th subcarrier, the $l$th AP multiplies the conjugate of channel estimate (perfect channel state information) by the received signal \eqref{frequency}. Then the derived quantity ${{\overset{\lower0.5em\hbox{$\smash{\scriptscriptstyle\smile}$}}{y} }_{kl}}\left[ s \right] = {\mathbf{h}}_{kl}^{\text{H}}\left[ s \right]{{\mathbf{y}}_l}\left[ s \right]$ is sent to the CPU through the fronthaul. The CPU uses the weight coefficients ${a_{kl}\left[ s \right]}$ to obtain ${{\hat y}_k}\left[ s \right]$ as
\begin{align}
  &{{\hat y}_k}\left[ s \right] = \sum\limits_{l = 1}^L {a_{kl}^*\left[ s \right]{{{{\overset{\lower0.5em\hbox{$\smash{\scriptscriptstyle\smile}$}}{y} }}}_{kl}}\left[ s \right]}  =  \sum\limits_{l = 1}^L {\underbrace {a_{kl}^*\left[ s \right]{\mathbf{h}}_{kl}^{\text{H}}\left[ s \right]}_{{\text{N}}{{\text{S}}_k}\left[ s \right]}} {{\mathbf{w}}_l}\left[ s \right] \notag\\
  &+ \underbrace {\sum\limits_{l = 1}^L {\sqrt {{p_k}} a_{kl}^*\left[ s \right]{I_{kl}}\left[ 0 \right]{\mathbf{h}}_{kl}^{\text{H}}\left[ s \right]{{\mathbf{h}}_{kl}}\left[ s \right]} }_{{\text{D}}{{\text{S}}_k}\left[ s \right]}{x_k}\left[ s \right] \notag \\
   &+ \sum\limits_{m \ne s}^M {\underbrace {\sum\limits_{l = 1}^L {\sqrt {{p_k}} a_{kl}^*\left[ s \right]{I_{kl}}\left[ {m - s} \right]{\mathbf{h}}_{kl}^{\text{H}}\left[ s \right]{{\mathbf{h}}_{kl}}\left[ m \right]} }_{{\text{IC}}{{\text{I}}_k}\left[ {m - s} \right]}} {x_k}\left[ m \right] \notag\\
   &+ \!\sum\limits_{i \ne k}^K {\!\sum\limits_{m = 1}^M {\underbrace {\!\sum\limits_{l = 1}^L {\!\sqrt {{p_i}} a_{kl}^*\left[ s \right]\!{I_{il}}\left[ {m - s} \right]\!{\mathbf{h}}_{kl}^{\text{H}}\left[ s \right]\!{{\mathbf{h}}_{il}}\left[ m \right]} }_{{\text{U}}{{\text{I}}_{ki}}\left[ {m - s} \right]}} } {x_i}\left[ m \right],
\end{align}
where ${{\text{D}}{{\text{S}}_k}\left[ s \right]}$ represents the desired signal at $s$th subcarrier, ${{\text{IC}}{{\text{I}}_k}\left[ {m - s} \right]}$ represents the intercarrier interference from $m$th subcarrier, ${{\text{U}}{{\text{I}}_{ki}}\left[ {m - s} \right]}$ denotes the interference resulted from transmitted data from other TAs, and ${{\text{N}}{{\text{S}}_k}\left[ s \right]}$ represents the noise term, respectively.
\begin{thm}\label{thm1}
The capacity of TA $k$ at $s$th subcarrier is \eqref{SINR_CF} at the top of next page, where
\setcounter{equation}{9}
\begin{align}
  {{\mathbf{a}}_k}\left[ s \right] &\triangleq {\left[ {{a_{k1}}\left[ s \right] \ldots {a_{kL}}\left[ s \right]} \right]^{\mathrm{T}}} \in {\mathbb{C}^L} \notag \\
  {{\mathbf{b}}_k} &\triangleq {\left[ {{I_{k1}}\left[ 0 \right]{\beta _{k1}} \ldots {I_{kL}}\left[ 0 \right]{\beta _{kL}}} \right]^{\mathrm{T}}} \in {\mathbb{C}^L} \notag \\
  {{\mathbf{c}}_k}\left[ {m \!-\! s} \right] &\triangleq {\left[ {{I_{k1}}\left[ {m \!-\! s} \right]{\beta _{k1}} \ldots {I_{kL}}\left[ {m \!-\! s} \right]{\beta _{kL}}} \right]^{\mathrm{T}}} \in {\mathbb{C}^L} \notag \\
  {{\mathbf{d}}_{ki}}\left[ {m \!-\! s} \right] &\triangleq \left[ {{I_{i1}}\left[ {m \!-\! s} \right]\sqrt {{\beta _{k1}}{\beta _{i1}}} \eta \left( {{\varphi _{k1}},{\varphi _{i1}}} \right)} \right. \notag \\
  &\;\;\;{\left. { \ldots {I_{iL}}\left[ {m \!-\! s} \right]\sqrt {{\beta _{kL}}{\beta _{iL}}} \eta \left( {{\varphi _{kL}},{\varphi _{iL}}} \right)} \right]^{\mathrm{T}}} \in {\mathbb{C}^L}  \notag \\
  {{\mathbf{\Lambda }}_k} &\triangleq {\mathrm{diag}}\left( {{\beta _{k1}}, \ldots ,{\beta _{kL}}} \right) \in {\mathbb{C}^{L \times L}} .
\end{align}
\end{thm}
\begin{IEEEproof}
Please refer to Appendix A.
\end{IEEEproof}
It is worth noting that the weight vector ${{\mathbf{a}}_k}\left[ s \right]$ is different for each $s$ and can be optimized at the CPU to maximize the SE with the help of the LSFD receiver cooperation from [3].
\begin{cor}
The effective SINR of TA $k$ at $s$th subcarrier is maximized by
\begin{align}
  &{{\mathbf{a}}_k}\left[ s \right] = \left( {{p_k}{N^2}\sum\limits_{m \ne s}^M {{{\mathbf{c}}_k}\left[ {m - s} \right]{\mathbf{c}}_k^{\mathrm{H}}\left[ {m - s} \right]} } \right. \notag \\
   &+ {\sum\limits_{i \ne k}^K {\left. {{p_i}\sum\limits_{m \ne s}^M {{{\mathbf{d}}_{ki}}\left[ {m \!-\! s} \right]{\mathbf{d}}_{ki}^{\mathrm{H}}\left[ {m \!-\! s} \right]}  \!+\! {\sigma ^2}N{{\mathbf{\Lambda }}_k}} \right)} ^{ - 1}}\!{{\mathbf{b}}_k},
\end{align}
which leads to the maximum SE in \eqref{SINR_LSFD} at the top of next page.
\end{cor}
If we want to reduce the complexity of LSFD, then the conventional MF receiver cooperation from \cite{Ngo2017Cell} is obtained by using equal weights ${{\mathbf{a}}_k}\left[ s \right]{\text{ = }}{\left[ {1/L \ldots 1/L} \right]^{\text{T}}}$.

\newcounter{mytempeqncnt}
\begin{figure*}[t!]
\normalsize
\setcounter{mytempeqncnt}{1}
\setcounter{equation}{8}
\begin{align}\label{SINR_CF}
{\text{S}}{{\text{E}}_k}\left[ s \right] = {\log _2}\left( {1 + \frac{{{p_k}{N^2}{{\left| {{\mathbf{a}}_k^{\text{H}}\left[ s \right]{{\mathbf{b}}_k}\left[ s \right]} \right|}^2}}}{{{p_k}{N^2}\sum\limits_{m \ne s}^M {{{\left| {{\mathbf{a}}_k^{\text{H}}\left[ s \right]{{\mathbf{c}}_k}\left[ {m - s} \right]} \right|}^2}}  + \sum\limits_{i \ne k}^K {{p_i}\sum\limits_{m \ne s}^M {{{\left| {{\mathbf{a}}_k^{\text{H}}\left[ s \right]{{\mathbf{d}}_{ki}}\left[ {m - s} \right]} \right|}^2}} }  + {\sigma ^2}N{\mathbf{a}}_k^{\text{H}}\left[ s \right]{{\mathbf{\Lambda }}_k}\left[ s \right]{{\mathbf{a}}_k}\left[ s \right]}}} \right).
\end{align}
\setcounter{equation}{7}
\hrulefill
\end{figure*}

\newcounter{mytempeqncnt1}
\begin{figure*}[t!]
\normalsize
\setcounter{mytempeqncnt}{1}
\setcounter{equation}{11}
\begin{align}\label{SINR_LSFD}
{\text{SE}}_k^{{\text{LSFD}}}\left[ s \right] \!=\! {\log _2}\left( {1 \!+\! {p_k}{N^2}{\mathbf{b}}_k^{\text{H}}{{\left( {{p_k}{N^2}\!\sum\limits_{m \ne s}^M {{{\mathbf{c}}_k}\left[ {m \!-\! s} \right]{\mathbf{c}}_k^{\text{H}}\left[ {m \!-\! s} \right]}  \!+\! \sum\limits_{i \ne k}^K {{p_i}\sum\limits_{m \ne s}^M {{{\mathbf{d}}_{ki}}\left[ {m \!-\! s} \right]{\mathbf{d}}_{ki}^{\text{H}}\left[ {m \!-\! s} \right]}  \!+\! {\sigma ^2}N{{\mathbf{\Lambda }}_k}} } \right)}^{ - 1}}\!{{\mathbf{b}}_k}} \right).
\end{align}
\setcounter{equation}{7}
\hrulefill
\end{figure*}

\newcounter{mytempeqncnt2}
\begin{figure*}[t!]
\normalsize
\setcounter{mytempeqncnt}{1}
\setcounter{equation}{13}
\begin{align}\label{SINR_SC}
{\text{S}}{{\text{E}}_k^{\text{small cell}}}\left[ s \right] = \mathop {\max }\limits_{l \in \left\{ {1, \ldots ,L} \right\}} {\log _2}\left( {1 + \frac{{{p_k}{N^2}\beta _{kl}^2{{\left| {{I_{kl}}\left[ 0 \right]} \right|}^2}}}{{{p_k}{N^2}\beta _{kl}^2\sum\limits_{m \ne s}^M {{{\left| {{I_{kl}}\left[ {m - s} \right]} \right|}^2} + \sum\limits_{i \ne k}^K {{p_i}} {\beta _{kl}}{\beta _{il}}\mu \left( {{\varphi _{kl}},{\varphi _{il}}} \right) + {\sigma ^2}N{\beta _{kl}}} }}} \right).
\end{align}
\setcounter{equation}{7}
\hrulefill
\end{figure*}

\newcounter{mytempeqncnt3}
\begin{figure*}[t!]
\normalsize
\setcounter{mytempeqncnt}{1}
\setcounter{equation}{17}
\begin{align}\label{SINR_cellular}
{\text{S}}{{\text{E}}_k^\text{c}}\left[ s \right] = {\log _2}\left( {1 + \frac{{{p_k}{L^2}{N^2}\beta _k^2{{\left| {{I_k}\left[ 0 \right]} \right|}^2}}}{{{p_k}{L^2}{N^2}\beta _k^2\sum\limits_{m \ne s}^M {{{\left| {{I_k}\left[ {m - s} \right]} \right|}^2} + \sum\limits_{i \ne k}^K {{p_i}} {\beta _k}{\beta _i}\mu \left( {{\varphi _k},{\varphi _i}} \right) + {\sigma ^2}LN{\beta _k}} }}} \right).
\end{align}
\setcounter{equation}{12}
\hrulefill
\end{figure*}

\subsection{Small Cell With OFDM}

In this section, we compare with a conventional small cell system that consists of $L$ APs and $K$ TAs. The APs and TAs are at the same locations as in the CF case but each TA is only served by one AP, which gives the highest SE performance. The perfect channel estimate of TA $k$ is used to multiply the received signal for detecting the desired signal. So we can get the combined uplink signal at the $l$th AP is
\begin{align}
  &{{\overset{\lower0.5em\hbox{$\smash{\scriptscriptstyle\smile}$}}{y} }_{kl}}\left[ s \right] = {\mathbf{h}}_{kl}^{\text{H}}\left[ s \right]{{\mathbf{y}}_l}\left[ s \right] =  \underbrace {\sqrt {{p_k}} {I_{kl}}\left[ 0 \right]{\mathbf{h}}_{kl}^{\text{H}}\left[ s \right]{{\mathbf{h}}_{kl}}\left[ s \right]}_{{I_{kl,1}}\left[ s \right]}{x_k}\left[ s \right] \notag \\
   & \!+\! \sum\limits_{m \ne s}^M {\underbrace {\sqrt {{p_k}} {I_{kl}}\left[ {m - s} \right]{\mathbf{h}}_{kl}^{\text{H}}\left[ s \right]{{\mathbf{h}}_{kl}}\left[ m \right]}_{{I_{kl,2}}\left[ {m - s} \right]}{x_k}\left[ m \right]} \!+\! {\mathbf{h}}_{kl}^{\text{H}}\left[ s \right]{{\mathbf{w}}_l}\left[ s \right]  \notag \\
   &+ \sum\limits_{i \ne k}^K {\sum\limits_{m = 1}^M {\underbrace {\sqrt {{p_i}} {I_{il}}\left[ {m - s} \right]{\mathbf{h}}_{kl}^{\text{H}}\left[ s \right]{{\mathbf{h}}_{il}}\left[ m \right]}_{{I_{kil,3}}\left[ {m - s} \right]}{x_i}\left[ m \right]} } ,
\end{align}
where ${{I_{kl,1}}\!\left[ s \right]}$ is the desired signal at the $s$th subcarrier, ${{I_{kl,2}}\!\left[ {m \!-\! s} \right]}$ is the intercarrier interference from the $m$th subcarrier, and ${{I_{kil,3}}\!\left[ {m \!-\! s} \right]}$ is the interference caused by transmitted data from other TAs.
\begin{thm}
The small cell system is a special case of CF massive MIMO when the LSFD weights are selected so that each user is only served by the SE-maximizing AP.
Following similar steps in Theorem \ref{thm1} for CF massive MIMO-OFDM systems, the capacity of TA $k$ is given by \eqref{SINR_SC} at the top of next page, where $\mu \left( {{\varphi _{kl}},{\varphi _{il}}} \right) = N^2$ as ${\sin \left( {{\varphi _{il}}} \right) = \sin \left( {{\varphi _{kl}}} \right)}$. While ${\sin \left( {{\varphi _{il}}} \right) \ne \sin \left( {{\varphi _{kl}}} \right)}$, we have
\setcounter{equation}{14}
\begin{align}
\mu \left( {{\varphi _{kl}},{\varphi _{il}}} \right) = \frac{{{{\sin }^2}\left( {\pi {d_\mathrm{H}}N\left( {\sin \left( {{\varphi _{il}}} \right) - \sin \left( {{\varphi _{kl}}} \right)} \right)} \right)}}{{{{\sin }^2}\left( {\pi {d_\mathrm{H}}\left( {\sin \left( {{\varphi _{il}}} \right) - \sin \left( {{\varphi _{kl}}} \right)} \right)} \right)}}.
\end{align}
\end{thm}
\begin{IEEEproof}
Please refer to Appendix B.
\end{IEEEproof}
\subsection{Cellular Massive MIMO-OFDM}

We consider a cellular network with one cell and $LN$ antennas at the cellular BS. The channel gain between BS and TA $k$ at the $n$th time interval can be modeled as
\begin{align}
  {\mathbf{h}}_k^{\text{c}}\left[ n \right] &= \sqrt {{\beta _k}\left[ n \right]} \left[ {1\;\exp \left( {j2\pi {d_\text{H}}\sin \left( {{\varphi _k}\left[ n \right]} \right)} \right) \ldots } \right. \notag \\
  &{\left. {\exp \left( {j2\pi {d_\text{H}}(LN - 1)\sin \left( {{\varphi _k}\left[ n \right]} \right)} \right)} \right]^\text{T}} ,
\end{align}
where ${\beta _k}\left[ n \right]$ is the path loss between BS and TA $k$.
The perfect channel estimate of TA $k$ is used to multiply the received signal for detecting the desired signal. Then, the combined uplink signal at the BS is
\begin{align}
  &y_k^{\text{c}}\left[ s \right] = {\left( {{\mathbf{h}}_k^{\text{c}}\left[ s \right]} \right)^{\text{H}}}{{\mathbf{y}}^{\text{c}}}\left[ s \right] = \underbrace {\sqrt {{p_k}} {I_k}\left[ 0 \right]{{\left( {{\mathbf{h}}_k^{\text{c}}\left[ s \right]} \right)}^{\text{H}}}{\mathbf{h}}_k^{\text{c}}\left[ s \right]}_{{\Gamma _{k,1}}\left[ s \right]}{x_k}\left[ s \right] \notag \\
   &+ \!\sum\limits_{m \ne s}^M {\underbrace {\sqrt {{p_k}} {I_k}\left[ {m \!-\! s} \right]{{\left( {{\mathbf{h}}_k^{\text{c}}\left[ s \right]} \right)}^{\text{H}}}{\mathbf{h}}_k^{\text{c}}\left[ m \right]}_{{\Gamma _{k,2}}\left[ {m - s} \right]}{x_k}\left[ m \right]}  \!+\! {\left( {{\mathbf{h}}_k^{\text{c}}\left[ s \right]} \right)^{\text{H}}}{\mathbf{w}}\left[ s \right] \notag \\
   &+ \sum\limits_{i \ne k}^K {\sum\limits_{m = 1}^M {\underbrace {\sqrt {{p_i}} {I_i}\left[ {m - s} \right]{{\left( {{\mathbf{h}}_k^{\text{c}}\left[ s \right]} \right)}^{\text{H}}}{\mathbf{h}}_i^{\text{c}}\left[ m \right]}_{{\Gamma _{ki,3}}\left[ {m - s} \right]}{x_i}} } \left[ m \right] ,
\end{align}
where ${{\Gamma _{k,1}}\left[ s \right]}$, ${{\Gamma _{k,2}}\left[ {m - s} \right]}$ and ${{\Gamma _{ki,3}}\left[ {m - s} \right]}$ are the desired signal at the $s$th subcarrier, the intercarrier interference from the $m$th subcarrier and the interference caused by transmitted data from other TAs, respectively.
\begin{thm}
Using the maximum-ratio combining, the SE of TA $k$ is given by \eqref{SINR_cellular} at the top of this page, where $\mu \left( {{\varphi _k},{\varphi _i}} \right) = L^2N^2$ while $\sin \left( {{\varphi _i}} \right) = \sin \left( {{\varphi _k}} \right)$. When $\sin \left( {{\varphi _i}} \right) \ne \sin \left( {{\varphi _k}} \right)$, we have
\setcounter{equation}{18}
\begin{align}
\mu \left( {{\varphi _k},{\varphi _i}} \right) = {\frac{{{{\sin }^2}\left( {\pi {d_\mathrm{H}}LN\left( {\sin \left( {{\varphi _i}} \right) - \sin \left( {{\varphi _k}} \right)} \right)} \right)}}{{{{\sin }^2}\left( {\pi {d_\mathrm{H}}\left( {\sin \left( {{\varphi _i}} \right) - \sin \left( {{\varphi _k}} \right)} \right)} \right)}}}.
\end{align}
\end{thm}
\begin{IEEEproof}
It follows similar steps in Theorem \ref{thm1} for CF massive MIMO-OFDM systems.
\end{IEEEproof}

\section{Numerical Results and Discussion}\label{se:numerical}

\begin{table*}[t]
\centering
\newcommand{\tabincell}[2]{\begin{tabular}{@{}#1@{}}#2\end{tabular}}
\caption{The SE drop percentage of HST with CF, small cell and cellular massive MIMO-OFDM systems.}
\setlength{\tabcolsep}{4mm}{
\vspace{4mm}
\centering
\begin{tabular}{|c|c|c|c|c|c|}
    \hline
    \hline
     Network Scheme & Number of APs & Vertical Distance & Largest SE value & Smallest SE value & SE drop percentage \cr\hline
     CF (MF) & $L=20$ & $d_{\text{ve}} = 50$ m  & $\dot a= 0.97$ & $\ddot a=0.94$ & $\eta=3\%$ \cr\hline
     Small cell & $L=20$ & $d_{\text{ve}} = 50$ m  & $\dot a=1.39$ & $\ddot a=1.29$ & $\eta=7\%$ \cr\hline
     Cellular & $L=20$ & $d_{\text{ve}} = 50$ m  & $\dot a=4.06$ & $\ddot a=0.18$ & $\eta=96\%$ \cr\hline
     CF (LSFD) & $L=20$ & $d_{\text{ve}} = 50$ m  & $\dot a=2.76$ & $\ddot a=2.48$ & $\eta=10\%$ \cr\hline
     CF (LSFD) & $L=30$ & $d_{\text{ve}} = 50$ m  & $\dot a=3.83$ & $\ddot a=3.59$ & $\eta=6\%$ \cr\hline
     CF (LSFD) & $L=20$ & $d_{\text{ve}} = 200$ m  & $\dot a=0.84$ & $\ddot a=0.79$ & $\eta=6\%$ \cr\hline
    \hline
\end{tabular}}
\label{CSQ}
\end{table*}

We present simulation results to show the validity of our theoretical analysis in this section, and also provide practical insights on the ICI reduction.
We utilize a simulation setup where $L$ APs are independently and equally distributed on one side of the 1000 m high-speed railway, and $K$ TAs are equally distributed on the 200 m HST. We consider communication at the carrier frequency $f_c = 2$ GHz. All TAs have the transmitted power $p=200$ mW, the bandwidth is $B = 20$ MHz, the noise power is $\sigma^2 = -96$ dBm. In addition, we consider an OFDM system with sampling duration $T = 0.5$ ms, and the total subcarrier number $M = 64$.
Note that, we use ${\text{SE}} \!=\!\! \left(\! {\sum\nolimits_{k = 1}^K \!{\sum\nolimits_{s = 1}^M {{\text{S}}{{\text{E}}_k}\!\left[ s \right]} } } \right)\!/\!\left( {K\!M} \right)$ to evaluate performance, and Average SE is the average SE at all positions HST travels.
\begin{figure}[t]
\centering
\includegraphics[scale=0.55]{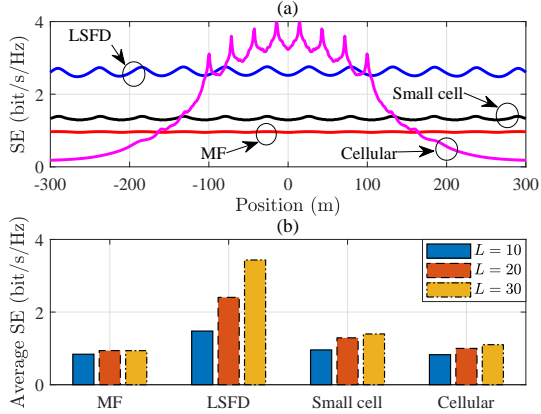}
\caption{(a) SE against the position of HST in CF, small cell and cellular massive MIMO-OFDM systems ($L\!=\!20$, $K\!=\!8$, $N\!=\!2$, $v\!=\!300\text{ km/h}$, $d_{\text{ve}}\!=\!50\text{ m}$). (b) Average SE for CF, small cell and cellular massive MIMO-OFDM systems ($K\!=\!8$, $N\!=\!2$, $v\!=\!300\text{ km/h}$, $d_{\text{ve}}\!=\!50\text{ m}$).} \vspace{-4mm}
\label{figure1}
\end{figure}

\begin{figure}[t]
\centering
\includegraphics[scale=0.55]{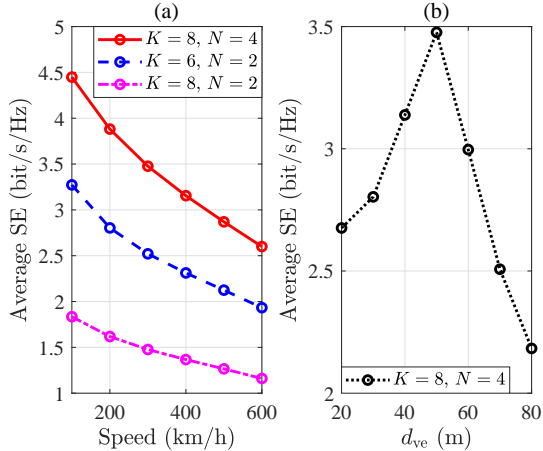}
\caption{Average SE of HST under different parameters in CF massive MIMO-OFDM systems with LSFD receiver ($L=10$). (a) $d_\text{ve}=50$ m; (b) $v=300$ km/h.} \vspace{-4mm}
\label{figure2}
\end{figure}
Fig.~\ref{figure1} (a) compares the SE against the position of HST in the CF (MF, LSFD), small cell and cellular massive MIMO-OFDM systems, respectively. We find that the CF massive MIMO-OFDM system with the MF receiver achieves a worse performance than the small cell system in HST communications. The reason is that the equal gain combing from APs to CPU can result in large inter TA interference and DFO interference in HST communications due to parallel structures among APs and TAs. However, CF massive MIMO-OFDM systems with LSFD can effectively reduce interference to get two-fold performance gain than small cell systems. Therefore, in HST communications, effective signal processing like LSFD is very necessary. Moreover, HST communications with cellular massive MIMO-OFDM systems also are considered for comparison. It is clear that a great SE gain can be achieved at the position near the BS, while it has poor SE performance in far positions.
In addition, Fig.~\ref{figure1} (b) shows the average SE for CF, small cell and cellular massive MIMO-OFDM systems with different number of APs, respectively. It is found that the LSFD receiver can achieve the largest average SE among all considered systems. Moreover, as the number of APs increases from $L=10$ to $L=30$, the LSFD receiver can obtain two-fold SE performance gain, but not the MF, small cell and cellular systems. The reason is that LSFD can not only obtain macro diversity gain, but also reduce interference.

The average SE against different speed of HST in CF massive MIMO-OFDM systems with LSFD receiver is shown in Fig.~\ref{figure2} (a). It is clear that the average SE decreases with the increasing of the speed of the HST, due to the DFO effect. It is also found that decreasing the number of TAs and increasing the number of antennas per AP can improve the SE performance.
In addition, Fig.~\ref{figure2} (b) shows the average SE against the vertical distance between APs and HST in the considered system. It is clear that the average SE first increases and then decreases with the increase of the vertical distance between APs and HST. There is an optimal $d_\text{ve}$ to achieves the maximum average SE. The reason is that the DFO effect decreases with the increase of the vertical distance between APs and HST, but the path loss increases quickly.


Furthermore, in $[-300 \text{ m},300 \text{ m}]$ moving range of HST, we select the largest SE value as $\dot a$ and the smallest SE value as $\ddot a$. Therefore, we define $\eta = \left(\dot a-\ddot a\right)/\dot a$ as the SE drop percentage to indicate the uniformity of system performance. Besides, the SE drop percentage for different considered systems are shown as TABLE~\ref{CSQ}. It is easy to find that cellular massive MIMO-OFDM systems has the maximal SE drop percentage in HST communications. Besides, CF massive MIMO-OFDM systems with LSFD receiver can not only achieve large SE, but also get low SE drop percentage. Moreover, increasing the number of APs $L$ and the vertical distance $d_{\text{ve}}$ both can reduce the SE drop percentage of HST communications.

\section{Conclusion}\label{se:conclusion}

In this paper, we investigate the uplink SE of CF massive MIMO-OFDM systems for HST communications with large ICI caused by DFO. We derive the closed-form expressions for uplink SE of CF massive MIMO systems with both MF and LSFD receivers and in order to quantify the DFO effect. In addition, HST communications with small cell and cellular massive MIMO systems are analysed for comparison. Compared with other considered systems, both high SE and low SE drop percentage can be achieved in LSFD receiver. Besides, more APs and antennas per AP can obviously improve the SE and reduce SE drop percentage, and there is an optimal vertical distance between APs and HST to balance DFO loss and path gain hence achieving the maximum SE.

\vspace{-0.3cm}
\begin{appendices}
\section{Proof of Theorem 1}

The SE of TA $k$ at the $s$th subcarrier is
\begin{align}\label{SE_k}
{\text{S}}{{\text{E}}_k}\left[ s \right] = {\log _2}\left( {1 + {\text{SIN}}{{\text{R}}_k}\left[ s \right]} \right),
\end{align}
with ${{\text{SIN}}{{\text{R}}_k}\left[ s \right]}$ is given as
\begin{align}
{\frac{{{{\left| {{\text{D}}{{\text{S}}_k}\left[ s \right]} \right|}^2}}}{{\sum\limits_{m \ne s}^M \!\!{{{\left| {{\text{IC}}{{\text{I}}_k}\!\left[ {m \!-\! s} \right]} \right|}^2}} \!+\!\! \sum\limits_{i \!\ne\! k}^K \!{\sum\limits_{m \!=\! 1}^M \!\!{{{\left| {{\text{U}}{{\text{I}}_{ki}}\!\left[ {m \!-\! s} \right]} \right|}^2}} }  \!+\! {\sigma ^2}\!\sum\limits_{l = 1}^L \!\!{{{\left\| {{\text{N}}{{\text{S}}_k}\left[ s \right]} \right\|}^2}} }}}.
\end{align}
Because the time-delay at each path has been ignored, we consider the MIMO channels as flat fading. Therefore, at different subcarriers, we have ${{\mathbf{h}}_{kl}}\left[ 1 \right] = {{\mathbf{h}}_{kl}}\left[ 2 \right] =  \ldots  = {{\mathbf{h}}_{kl}}\left[ M \right] = {{\mathbf{h}}_{kl}}$. With the help of \eqref{h_kl}, we can obtain
\begin{align}\label{h1}
{\mathbf{h}}_{kl}^{\text{H}}{{\mathbf{h}}_{kl}} = N{\beta _{kl}}.
\end{align}
For different TAs, with the help of the geometric series formula $\sum\nolimits_{m = 0}^{M - 1} {{x^m} = \left( {1 - {x^M}} \right)/\left( {1 - x} \right)}$ for $x \ne 1$ and $\sum\nolimits_{m = 0}^{M - 1} {{x^m} = M}$ for $x=1$, we have
\begin{align}\label{h2}
  {\mathbf{h}}_{kl}^{\text{H}}{{\mathbf{h}}_{il}} &\!=\! \sqrt {{\beta _{kl}}{\beta _{il}}}\!\sum\limits_{\lambda  = 0}^{N - 1} \!\!{{{\left( {\exp\! \left( {j2\pi {d_\text{H}}\!\left( {\sin\! \left( {{\varphi _{il}}} \right) \!-\! \sin\! \left( {{\varphi _{kl}}} \right)} \right)} \right)} \right)}^\lambda }}  \notag\\
  &=\! \sqrt {{\beta _{kl}}{\beta _{il}}} \eta \left( {{\varphi _{kl}},{\varphi _{il}}} \right),
\end{align}
where $\eta \left( {{\varphi _{kl}},{\varphi _{il}}} \right) \triangleq N$ when ${\sin \left( {{\varphi _{il}}} \right) = \sin \left( {{\varphi _{kl}}} \right)}$. As while as ${\sin \left( {{\varphi _{il}}} \right) \ne \sin \left( {{\varphi _{kl}}} \right)}$, we have
\begin{align}
\eta \left( {{\varphi _{kl}},{\varphi _{il}}} \right) \!\triangleq\! \frac{{1 \!-\! {\exp\! \left({j2\pi {d_\text{H}}N\left( {\sin \left( {{\varphi _{il}}} \right) \!-\! \sin \left( {{\varphi _{kl}}} \right)} \right)}\right)}}}{{1 \!-\! {\exp\!\left({j2\pi {d_\text{H}}\left( {\sin \left( {{\varphi _{il}}} \right) \!-\! \sin \left( {{\varphi _{kl}}} \right)} \right)}\right)}}}.
\end{align}
Using \eqref{h1} and \eqref{h2}, we can derive
\begin{align}
  \label{DS}{\left| {{\text{D}}{{\text{S}}_k}\left[ s \right]} \right|^2} &= {\left| {\sum\limits_{l = 1}^L {\sqrt {{p_k}} a_{kl}^*\left[ s \right]{I_{kl}}\left[ 0 \right]{\mathbf{h}}_{kl}^{\text{H}}\left[ s \right]{{\mathbf{h}}_{kl}}\left[ s \right]} } \right|^2} \notag \\
   &= {p_k}{N^2}{\left| {\sum\limits_{l = 1}^L {a_{kl}^*\left[ s \right]{I_{kl}}\left[ 0 \right]{\beta _{kl}}} } \right|^2}, \\
  \label{ICI}{\left| {{\text{IC}}{{\text{I}}_k}\left[ {m \!-\! s} \right]} \right|^2} &\!=\! {\left| {\sum\limits_{l = 1}^L {\sqrt {{p_k}} a_{kl}^*\left[ s \right]\!{I_{kl}}\left[ {m \!-\! s} \right]\!{\mathbf{h}}_{kl}^{\text{H}}\left[ s \right]\!{{\mathbf{h}}_{kl}}\left[ m \right]} } \right|^2} \notag \\
   &= {p_k}{N^2}{\left| {\sum\limits_{l = 1}^L {a_{kl}^*\left[ s \right]{I_{kl}}\left[ {m - s} \right]{\beta _{kl}}} } \right|^2}, \\
  \label{UI}{\left| {{\text{U}}{{\text{I}}_{ki}}\left[ {m \!-\! s} \right]} \right|^2} &\!=\! {\left| {\sum\limits_{l = 1}^L {\sqrt {{p_i}} a_{kl}^*\left[ s \right]{I_{il}}\left[ {m \!-\! s} \right]{\mathbf{h}}_{kl}^{\text{H}}\left[ s \right]{{\mathbf{h}}_{il}}\left[ m \right]} } \right|^2} \notag \\
   &\!\!\!\!\!\!\!\!\!\!\!\!\!\!\!\!\!\!\!\!\!=\! {p_i}{\left| {\sum\limits_{l = 1}^L {a_{kl}^*\left[ s \right]{I_{il}}\left[ {m \!-\! s} \right]\!\sqrt {{\beta _{kl}}{\beta _{il}}} \eta \left( {{\varphi _{kl}},{\varphi _{il}}} \right)} } \right|^2}, \\
   \label{NS}{\left\| {{\text{N}}{{\text{S}}_k}\left[ s \right]} \right\|^2} &\!=\! {\left\| {a_{kl}^*\left[ s \right]{\mathbf{h}}_{kl}^{\text{H}}\left[ s \right]} \right\|^2} \!=\! N\!\sum\limits_{l = 1}^L {{{\left| {a_{kl}^*\left[ s \right]} \right|}^2}{\beta _{kl}}}.
\end{align}
Finally, submitting \eqref{DS}, \eqref{ICI}, \eqref{UI} and \eqref{NS} into \eqref{SE_k} to finish the proof.
\section{Proof of Theorem 2}

By the Euler's formulas, we have
\begin{align}
  &{\left| {\eta \left( {{\varphi _{kl}},{\varphi _{il}}} \right)} \right|^2} \!=\! {\left|\! {\frac{{1 \!-\! \exp\! \left( {j2\pi {d_\text{H}}N\!\left( {\sin\! \left( {{\varphi _{il}}} \right) \!-\! \sin\! \left( {{\varphi _{kl}}} \right)} \right)} \right)}}{{1 \!-\! \exp\! \left( {j2\pi {d_\text{H}}\left( {\sin\! \left( {{\varphi _{il}}} \right) \!-\! \sin\! \left( {{\varphi _{kl}}} \right)} \right)} \right)}}} \!\right|^2} \notag \\
   &= \left| {\frac{{\exp \left( {\pi j{d_\text{H}}N\left( {\sin \left( {{\varphi _{il}}} \right) - \sin \left( {{\varphi _{kl}}} \right)} \right)} \right)}}{{\exp \left( {\pi j{d_\text{H}}\left( {\sin \left( {{\varphi _{il}}} \right) - \sin \left( {{\varphi _{kl}}} \right)} \right)} \right)}}} \right. \notag \\
   &\times {\left. {\frac{{\sin \left( {\pi {d_\text{H}}N\left( {\sin \left( {{\varphi _{il}}} \right) - \sin \left( {{\varphi _{kl}}} \right)} \right)} \right)}}{{\sin \left( {\pi {d_\text{H}}\left( {\sin \left( {{\varphi _{il}}} \right) - \sin \left( {{\varphi _{kl}}} \right)} \right)} \right)}}} \right|^2} \notag \\
   &= \frac{{{{\sin }^2}\left( {\pi {d_\text{H}}N\left( {\sin \left( {{\varphi _{il}}} \right) \!-\! \sin \left( {{\varphi _{kl}}} \right)} \right)} \right)}}{{{{\sin }^2}\left( {\pi {d_\text{H}}\left( {\sin \left( {{\varphi _{il}}} \right) \!-\! \sin \left( {{\varphi _{kl}}} \right)} \right)} \right)}} \!=\! \mu \left( {{\varphi _{kl}},{\varphi _{il}}} \right) .
\end{align}
Finally, with the help of \eqref{I_kl}, we obtain
\begin{align}
\sum\nolimits_{m = 1}^M {{{\left| {{I_{kl}}\left[ {m - s} \right]} \right|}^2} = 1} ,\forall k,\forall l,\forall s
\end{align}
to finish the proof.

\end{appendices}

\vspace{0.3cm}
\bibliographystyle{IEEEtran}
\bibliography{IEEEabrv,Ref}

\end{document}